\documentclass[11pt, a4paper, prd, twocolumn, preprintnumbers]{revtex4}

\usepackage{psfig, epsfig,color,amsmath,amssymb, mathrsfs}
\usepackage{graphicx}
\usepackage{fancyhdr, url, amsfonts}

\newcommand{\be}[1]{\small \begin{equation} #1 \end{equation}}

\begin{document}

\title{Perturbations around black holes}

\author{Bin Wang}
\affiliation{Department of Physics, Fudan University, Shanghai
200433, China} \email{wangb@fudan.edu.cn}


\date{\today}

\begin{abstract}
Perturbations around black holes have been an intriguing topic in
the last few decades. They are particularly important today, since
they relate to the gravitational wave observations which may provide
the unique fingerprint of black holes' existence. Besides the
astrophysical interest, theoretically perturbations around black
holes can be used as  testing grounds to examine the proposed
AdS/CFT and dS/CFT correspondence.
\end{abstract}

\maketitle

\section{Introduction}
Present astronomical searching for black holes are based on the
study of images of strong X-rays taken of matter around the probable
black hole candidate. Usually the astronomers study the X-ray binary
systems, which consist of a visible star in close orbit around an
invisible companion star which may be a  black hole. The companion
star pulls gas away from the visible star. As this gas forms a
flattened disk, it swirls toward the companion. Friction caused by
collisions between the particles in the gas heats them to extreme
temperatures and they produce X-rays that flicker or vary in
intensity within a second.  Many bright X-ray binary sources have
been discovered in our galaxy and nearby galaxies. In about ten of
these systems, the rapid orbital velocity of the visible star
indicates that the unseen companion is a black hole. The X-rays in
these objects are produced by particles very close to the event
horizon. In less than a second after they give off their X-rays,
they disappear beyond the event horizon. In a very recent study, the
sharpest images ever taken of matter around the probable black hole
at the centre of our Galaxy have brought us within grasp of a
crucial test of general relativity ¡ª a picture of the black hole's
`point of no return' \cite{1}.  A supermassive black hole has been
proved existing in  the centre of our own Galaxy.

However this way of searching for black hole is not direct, to some
sense it relies on the evolution behavior of the visible companion.
Whether a black hole itself has a characteristic `sound', which can
tell us its existence,  is a question we want to ask. Performing
numerical studies of perturbations around black holes, it was found
that during a certain time interval the evolution of the initial
perturbation is dominated by damped single-frequency oscillation.
The frequencies and damping of these oscillations relate only to the
black hole parameters, not to initial perturbations. This kind of
perturbation which is damped quite rapidly and exists only in a limited 
time interval is referred to as the quasinormal modes. They will dominate
most processes involving perturbed black holes and carry a unique
fingerprint which would lead to the direct identification of the
black hole existence. Detection of these quasinormal modes is
expected to be realized through gravitational wave observations in
the near future. In order to extract as much information as possible
from gravitational wave signal, it is important that we understand
exactly how the quasinormal modes behave for the parameters of black
holes in different models. (see \cite{2}\cite{3} for a review and
references therein)

Besides the astronomical interest, the perturbations around black
holes have profound theoretical implications.  Motivated by the
discovery of the correspondence between physics in the Anti-de
Sitter (AdS) spacetime and conformal field theory (CFT) on its
boundary (AdS/CFT), the investigation of QNM in AdS spacetimes
became appealing in the past several years. It was argued that the
QNMs of AdS black holes have direct interpretation in term of the
dual CFT (for an extensive but not exhaustive list see
\cite{Chan_Mann}\cite{Horowitz-00}\cite{Wang-00}\cite{Wang-01}\cite{Cardoso,Berti-03}\cite{Konoplya-02,
Birmingham,Zhu_Wang_Mann_Abdalla,
  AdS_SeveralAuthors,Cardoso_Lemos,Cardoso_Konoplya_Lemos})
. In de Sitter (dS) space the relation between bulk dS spacetime and
the corresponding CFT at the past boundary and future boundary in
the framework of scalar perturbation spectrums has also been
discussed \cite{deSitter_1,deSitter_2}\cite{du1}\cite{du2}. A
quantitative support of the dS/CFT correspondence has been provided. More
recently the quasinormal modes have also been argued as a possible
way to detect extra dimensions \cite{11}.

The study of quasinormal modes has been an intriguing subject. In 
this review we  will restrict ourselves to the
discussion of non-rotating black holes. We will first go over
perturbations in asymptotically flat spacetimes. In the following
section, we will focus on the perturbations in AdS spacetimes and dS
spacetimes and show that quasinormal modes around black holes are
testing grounds of AdS/CFT and dS/CFT correspondence. We will
present our conclusions and outlook in the last part.

\section{Perturbations in asymptotically flat spacetimes}

A great deal of effort has been devoted to the study of the
quasinormal modes concerned with black holes immersed in an
asymptotically flat spacetime. The perturbations of Schwarzschild
and Reissner-Nordstrom (RN) black holes can be reduced to simple
wave equations which have been examined extensively
\cite{2}\cite{3}. However, for nonspherical black holes one has to
solve coupled wave equations for the radial part and angular part,
respectively. For this reason the nonspherical case has been studied
less thoroughly, although there has recently been progress along
these lines \cite{12}. In asymptotically flat black hole backgrounds
radiative dynamics always proceeds in the same three stages: initial
impulse, quasinormal ringing and inverse power-law relaxation.

Introducing small perturbation $h_{\mu\nu}$ on a static spherically
symmetric background metric, we have the perturbed metric with the
form
\begin{equation}
  g_{\mu\nu} = g_{\mu\nu}^0 + h_{\mu\nu}.
  \label{scw_pert}
\end{equation}
In vacuum, the perturbed field equations simply reduce to
\begin{equation}
\delta R_{\mu\nu}=0.
\end{equation}
These equations are in linear in $h$.

For the spherically symmetric background, the perturbation is forced
to be considered with complete angular dependence. From the 10
independent components of the $h_{\mu\nu}$ only $h_{tt}$, $h_{tr}$,
and $h_{rr}$ transform as scalars under rotations. The $h_{t
\theta}$, $h_{t\phi}$, $h_{r\theta}$, and $h_{r\phi}$ transform as
components of two-vectors under rotations and can be expanded in a
series of vector spherical harmonics while the components
$h_{\theta\theta}$, $h_{\theta\phi}$, and $h_{\phi\phi}$ transform
as components of a $2\times2$ tensor and can be expanded in a series
of tensor spherical harmonics \cite{2}\cite{3}. There are two
classes of tensor spherical harmonics (polar and axial). The
differences are their parity under space inversion $(\theta,
\phi)\rightarrow (\pi -\theta, \pi +\phi)$.  After the inversion,
for the function acquiring a factor $(-1)^l$  refers to polar
perturbation, and the function acquiring a factor $(-1)^{l+1}$ is
called the axial perturbation.

The radial component of a perturbation outside the event horizon
satisfies the following wave equation,
\begin{equation}
  {{\partial^2}\over {\partial t^2}} \chi_\ell +
  \left( -{{\partial^2}\over {\partial r_*^2}} + V_\ell(r)
  \right)\chi_\ell = 0,
  \label{qnmwave}
\end{equation}
where $r_*$ is the ``tortoise'' radial coordinate. This equation
keeps the same form for both the axial and polar perturbations. The
difference between the axial and polar perturbations exists in their
effective potentials. For the axial perturbation around a
Schwarzschild black hole, the effective potential reads
\begin{equation}
  V_\ell(r)=\left(1 - {2M\over r}\right) \left[ {{\ell(\ell+1)}\over
  r^2} + {{2\sigma M} \over r^3} \right].
  \label{rwpot}
\end{equation}
However for the polar perturbation, the effective potential has the
form
\begin{widetext}
\be{ \label{zerpot}
\begin{split}
  V_\ell(r)=\left( 1 - \frac{2M}r \right) \frac{2n^2(n+1)r^3 + 6n^2Mr^2 +
  18nM^2r +18M^3}{r^3(nr+3M)^2}.
\end{split}
}
\end{widetext}
Apparently these two effective potentials look quite different,
however if we compare them numerically, we will find that they
exhibit nearly the same potential barrier outside the black hole
horizon, especially with the increase of $l$. Thus polar and axial
perturbations will give us the same quasinormal modes around the
black hole.

Solving the radial perturbation equation here is very similar to
solving the Schrodinger equation in quantum mechanics. We have a
potential barrier outside the black hole horizon, and the incoming
wave will be transmitted and reflected by this barrier. Thus many
methods developed in quantum mechanics can be employed here. In the
following we list some main results of quasinormal modes in
asymptotically flat spacetimes obtained before.

(a) It was found that all perturbations around black holes experience
damping behavior. This is interesting, since it tells us that black
hole solutions are stable.

(b) The quasinormal modes in black holes are isospectral. The same
quasinormal frequencies are found for different perturbations for
example axial and polar perturbations. This is due to the uniqueness
in which black holes react to perturbations.

(c) The damping time scale of the perturbation is proportional to
the  black hole mass, and it is shorter for higher-order modes
($\omega_{i,n+1} > \omega_{i,n}$). Thus the detection of
gravitational wave emitted from a perturbed black hole can be used
to directly measure  the black hole mass.

(d) Some properties of quasinormal frequencies can be learned from
the following table
\begin{table}[hptb]
  \begin{center}
    \begin{tabular}{|l|ll|ll|ll|}
      \hline
      n & $\ell=2$ & & $\ell=3$ & & $\ell=4$ & \cr
      \hline
      0 &  0.37367 & -0.08896 i & 0.59944 & -0.09270 i & 0.80918 &
      -0.09416 i \\
      1 &  0.34671 & -0.27391 i & 0.58264 & -0.28130 i & 0.79663 &
      -0.28443 i \\
      2 &  0.30105 & -0.47828 i & 0.55168 & -0.47909 i & 0.77271 &
      -0.47991 i \\
      3 &  0.25150 & -0.70514 i & 0.51196 & -0.69034 i & 0.73984 &
      -0.68392 i \\
      \hline
    \end{tabular}

\end{center}
  \caption{\it The first four QNM frequencies ($\omega M$) of the
    Schwarzschild black hole for $\ell =2, 3$, and
    $4$\cite{3}. }
  \label{table1}
\end{table}

We learned that for the same $l$, with the increase of $n$, the real
part of quasinormal frequencies decreases, while the imaginary part
increases. This corresponds to say that for the higher modes, the
perturbation will have less oscillations outside of the black hole
and die out quicker. With the increase of the multipole index $l$,
we found that both real part and imaginary part of quasinormal
frequencies increase, which shows that for the bigger $l$ the
perturbation outside the black hole will oscillate more but die out
quicker in the asymptotically flat spacetimes. This property will
change if one studies the AdS and dS spacetimes, since the behavior
of the effective potential will be changed there.

All previous works on quasinormal modes have so far been restricted
to time-independent black hole backgrounds. It should be realized
that, for a realistic model, the black hole parameters change with
time. A black hole gaining or losing mass via absorption (merging)
or evaporation is a good example. The more intriguing investigation
of the black hole quasinormal modes calls for a systematic analysis
of time-dependent spacetimes. Recently the late time tails under the
influence of a time-dependent scattering potential has been explored
in \cite{13}, where the tail structure was found to be modified due
to the temporal dependence of the potential. The
 exploration on the modification to the quasinormal modes
in time-dependent spacetimes has also been started. Instead of
plotting an effective time-dependent scattering potential by hand as
done in \cite{13}, we have introduced the time-dependent potential
in a natural way by considering dynamical black holes, with black
hole parameters changing with time due to absorption and evaporation
processes. We have studied the temporal evolution of massless scalar
field perturbation \cite{14}\cite{15}.

We found that the modification to the QNMs due to the time-dependent
background is clear. When the black hole mass $M$ increases linearly
with time, the decay becomes slower compared to the stationary case,
which corresponds to saying that $\vert\omega_i\vert$ decreases with
respect to time. The oscillation period is no longer a constant as
in the stationary Schwarzschild black hole. It becomes longer with
the increase of time. In other words, the real part of the
quasinormal frequency $\omega_r$ decreases with the increase of
time. When $M$ decreases linearly with respect to time, compared to
the stationary Schwarzschild black hole, we have observed that the
decay becomes faster and the oscillation period becomes shorter,
thus both $\vert\omega_i\vert$ and $\omega_r$ increase with time.
The objective picture can be seen in Fig.1.

\begin{figure}
\resizebox{1\linewidth}{!}{\includegraphics*{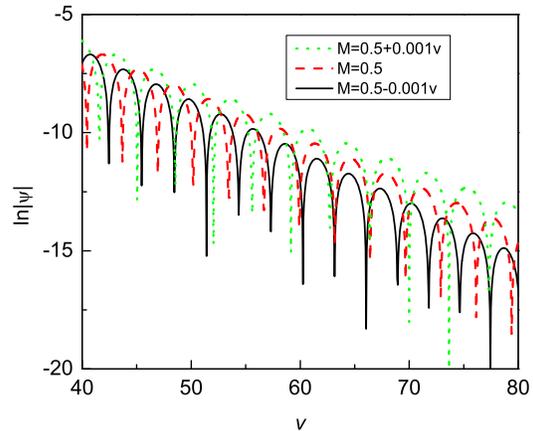}}
\caption{Temporal evolution of the field in the background of Vaidya
metric ($q$=0) for $l = 2$, evaluated at $r = 5$. The mass of the
black hole is $M(v) = 0.5\pm 0.001v$. The field evolution for $M(v)
= 0.5 + 0.001v$ and $M(v) = 0.5 - 0.001v$ are shown as the top curve
and the bottom curve respectively. For comparison, the oscillations
for $M = 0.5$ is given in the middle line.} \label{fig1}
\end{figure}

\section{Perturbations in AdS and dS black hole spacetimes}

Motivated by the recent discovery of the Anti-de Sitter/Conformal
Field Theory (AdS/CFT) correspondence, the investigation of QNMs in
Anti-de Sitter (AdS) spacetimes became appealing in the past years.
It was argued that the QNMs of AdS black holes have direct
interpretation in terms of the dual conformal field theory.

The first study of the QNMs in AdS spaces was performed by Chan and
Mann \cite{Chan_Mann}. Subsequently, Horowitz and Hubeny suggested a
numerical method to calculate the QN frequencies directly and made a
systematic investigation of QNMs for scalar perturbation on the
background of Schwarzschild AdS (SAdS) black holes
\cite{Horowitz-00}. They claimed that for large AdS black holes both
the real and imaginary parts of the quasinormal frequencies scale
linearly with the black hole temperature. However for small AdS
black holes they found a departure from this behaviour. This was
further confirmed by the object picture obtained in
\cite{Zhu_Wang_Mann_Abdalla}.

Considering that the Reissner-Nordstrom AdS (RNAdS) black hole
solution provides a better framework than the SAdS geometry and may
contribute significantly to our understanding of space and time, the
Horowitz-Hubeny numerical method was generalized to the study of
QNMs of RNAdS black holes in \cite{Wang-00} and later crosschecked
by using the time evolution approach \cite{Wang-01}.  Unlike the
SAdS case, the quasinormal frequencies do not scale linearly with
the black hole temperature, and the approach to thermal equilibrium
in the CFT was more rapid as the charge on the black hole increased.
In addition to the scalar perturbation, gravitational and
electromagnetic perturbations in AdS black holes have also attracted
attention \cite{Cardoso,Berti-03}. Other works on QNMs in AdS
spacetimes can be found in \cite{Konoplya-02,
Birmingham,Zhu_Wang_Mann_Abdalla,
  AdS_SeveralAuthors,Cardoso_Lemos,Cardoso_Konoplya_Lemos}.  Recently in \cite{Berti-03} Berti and Kokkotas used the frequency-domain
method and restudied the scalar perturbation in RNAdS black holes.
They verified most of our previous numerical results in
\cite{Wang-00,Wang-01}.

As was pointed out in \cite{Wang-00} and later supported in
\cite{Berti-03}, the Horowitz-Hubeny method breaks down for large
values of the charge. To study the QNMs in the near extreme and
extreme RNAdS backgrounds, we need to count on time evolution
approach. Employing an improved numerical method, we have shown that
the problem with minor instabilities in the form of ``plateaus'',
which were observed in \cite{Wang-01}, can be overcome. We obtained the
precise QNMs behavior in the highly charged RNAdS black holes.

To illustrate the properties of quasinormal modes in AdS black
holes, we here briefly review the perturbations around RNAdS black
holes. The metric describing a charged, asymptotically Anti-de
Sitter spherical black hole, written in spherical coordinates, is
given by
\begin{equation}
ds^{2} = -h(r) dt^{2} + h(r)^{-1} dr^{2} +
r^{2}(d\theta^{2}+\sin^{2}\theta d\phi^{2}) \ , \label{metric}
\end{equation}
where the function $h(r)$ is
\begin{equation}
h(r) = 1 - \frac{2m}{r} + \frac{Q^{2}}{r^{2}} - \frac{\Lambda
r^{2}}{3} \ .
\end{equation}
We are assuming a negative cosmological constant, usually written as
$\Lambda = -3/R^{2}$. The integration constants $m$ and $Q$ are the
black hole mass and electric charge respectively. The extreme value
of the black hole charge, $Q_{max}$, is given by the function of the
event horizon radius in the form $Q_{max}^2=r_+^2(1+3r_+^2/R^2)$.
Consider now a scalar perturbation field $\Phi$ obeying the massless
Klein-Gordon equation

\begin{equation}\label{boxphiequalzero}
\Box \Phi = 0  .
\end{equation}
The usual separation of variables in terms of a radial field and a
spherical harmonic $\textrm{Y}_{\ell,m}(\theta,\varphi)$,
\begin{equation}
\Phi=\sum_{\ell\,m} \frac{1}{r} \Psi(t,r)\textrm{Y}_{\ell
m}(\theta,\phi)  \ , \label{Ansazs_field}
\end{equation}
leads to Schr\"{o}dinger-type equations in the tortoise coordinate
for each value of $\ell$. Introducing the tortoise coordinate $r^*$ by 
$\frac{dr^*}{dr} = h(r)^{-1}$ and the null coordinates $u = t -
r^*$ and $v =  t + r^*$, the field equation can be written as
\begin{equation}
- 4 \frac{\partial^{2}\Psi}{\partial u \partial v}
 = V(r) \Psi \ ,
\label{u_v_wave_equation}
\end{equation}
where the effective potential $V$ is
\begin{equation}
V(r)=h(r)\left[\frac{\ell(\ell + 1)}{r^2}+\frac{2m}{r^{3}} -
  \frac{2Q^2}{r^{4}} + \frac{2}{R^{2}}\right] \ .
\end{equation}
Wave equation (\ref{u_v_wave_equation}) is useful to study the time
evolution of the scalar perturbation, in the context of an initial
characteristic value problem.

In terms of the ingoing Eddington coordinates $(v,r)$ and separating
the scalar field in a product form as
\begin{equation}
\Phi= \sum_{\ell m} \frac{1}{r} \psi(r) \textrm{Y}_{\ell
m}(\theta,\phi) e^{-i\omega v},
\end{equation}
the minimally-coupled scalar wave equation (\ref{boxphiequalzero})
may thereby be reduced to
\begin{equation}
h(r)\frac{\partial^{2}\psi (r)}{\partial r^{2}}
 + \left[ h'(r) - 2i\omega \right]
 \frac{\partial\psi (r)}{\partial r}-\tilde{V}(r) \psi(r) = 0 \,\, ,
\label{r_wave_equation}
\end{equation}
where $\tilde{V}(r) = V(r)/h(r) = h'(r)/r+\ell(\ell+1)/r^2$.

Introducing $x=1/r$, Eq.(\ref{r_wave_equation}) can be re-expressed
as Eqs.(15-18) in \cite{Wang-00}. These equations are appropriate to
directly obtain the QN frequencies using the Horowitz-Hubeny method.

We have used two different numerical methods to solve the wave
equations. The first method is the Horowitz-Hubeny method. The
second numerical methods we have employed is the discretization for
equation (\ref{u_v_wave_equation}) in the form
\begin{widetext}
\be{
\begin{split}
\left[ 1 - \frac{\Delta^{2}}{16}V(S)\right] \psi(N) =
\psi(E)+\psi(W)-\psi(S)-\frac{\Delta^{2}}{16}
       \left[V(S)\psi(S)+V(E)\psi(E)+V(W)\psi(W)\right] \,\,.
\end{split}
}
\end{widetext}
The points $N$, $S$, $W$ and $E$ are defined as usual: $N = (u +
\Delta, v + \Delta)$, $W = (u + \Delta, v)$, $E = (u, v + \Delta)$
and $S = (u,v)$. The local truncation error is of the order of
$O(\Delta^{4})$.

Figure 2 demonstrates the behaviors of the field with the increase
of the charge in RN AdS black hole background. Since the imaginary
and real parts of the quasinormal frequencies relate to the damping
time scale ($\tau_1=1/\omega_i$) and oscillation time scale
($\tau_2=1/\omega_r$), respectively. We learned that as $Q$
increases $\omega_i$ increases as well, which corresponds to the
decrease of the damping time scale. According to the AdS/CFT
correspondence, this means that for bigger $Q$, it is quicker for
the quasinormal ringing to settle down to  thermal equilibrium. From
figure 2 it is easy to see this property.  Besides, figure 2 also
tells us that the bigger $Q$ is, the lower frequencies of oscillation
will be. If we perturb a RN AdS black hole with high charge, the
surrounding geometry will not ``ring'' as much and as long as that
of the black hole with small $Q$. It is easy for the perturbation on
the highly charged AdS black hole background to return to thermal
equilibrium.
\begin{figure}
\resizebox{1\linewidth}{!}{\includegraphics*{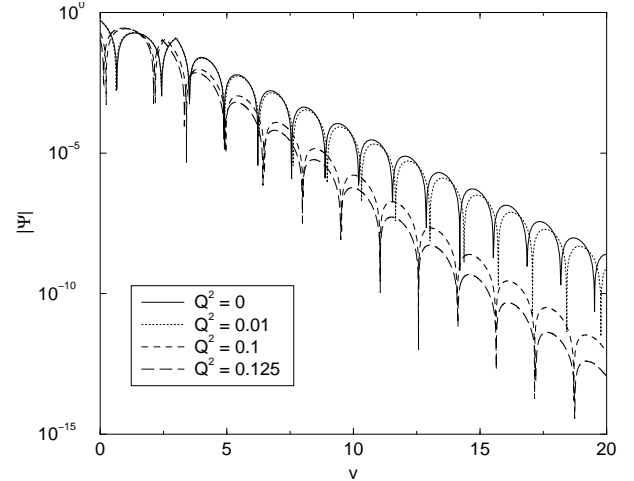}}
\caption{\it \small Semi--log graphs of $|\Psi|$ with $r_+ = 0.4$
and small values of $Q$. The extreme value for $Q^2$ is 0.2368.}
\label{fig2}
\end{figure}

However this relation seems  not to hold well when the charge is
sufficiently big. We see that over some critical value of $Q$, the
damping time scale increases with the increase of $Q$, corresponding
to the decrease of imaginary frequency. This means that over some
critical value of $Q$, the larger the black hole charge is, the
slower for the outside perturbation to die out. Besides the
oscillation starts to disappear at $Q_c=0.3895 Q_{max}$. These
points can be directly seen in the wave function plotting from
Fig.\ref{fig2} and Fig.\ref{fig4}.
\begin{figure}
\resizebox{1\linewidth}{!}{\includegraphics*{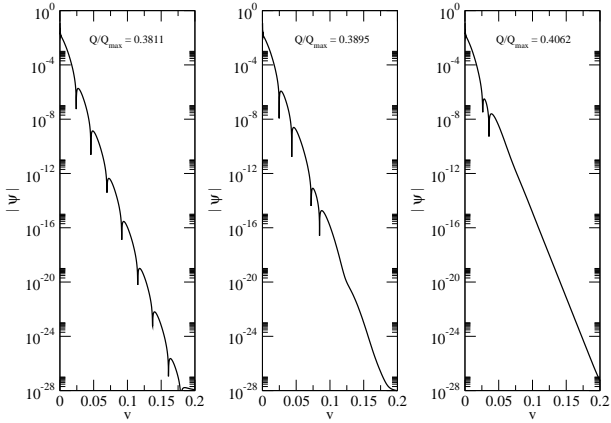}}
\caption{Semi-log graphs of the scalar field in the event horizon,
  showing the transition from oscillatory to non-oscillatory
  asymptotic decay. In the graphs, $r_{+}=100$, $\ell=0$ and $R=1$.}
\label{fig2}
\end{figure}
\begin{figure}
\resizebox{1\linewidth}{!}{\includegraphics*{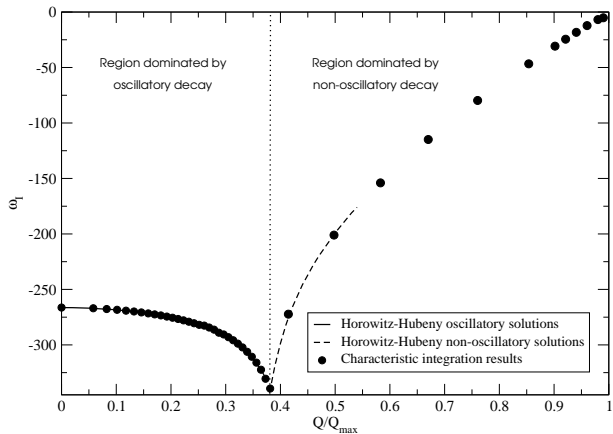}}
\caption{Graph of  $\omega_{i}$ with $Q/Q_{max}$, showing that
  $\omega_{i}$ tends to zero as $Q$ tends to $Q_{max}$. In the
  graph, $r_{+}=100$, $R=1$, $\ell=0$ and $n=0$.}
\label{fig4}
\end{figure}

In addition to the study of the lowest lying QNMs, it is interesting
to study the higher overtone QN frequencies for scalar
perturbations. The first attempt was carried out in
\cite{Cardoso_Konoplya_Lemos}. We have extended the study to the
RNAdS backgrounds.  It was argued that the dependence of the QN
frequencies on the angular index $\ell$ is extremely weak
\cite{Berti-03}. This was also claimed in
\cite{Cardoso_Konoplya_Lemos}. Using our numerical results we have
shown that this weak dependence on the angular index is not trivial.

For the same value of the charge, we have found that both real and
imaginary parts of QN frequencies increase with the overtone number
$n$, which is different from that observed in asymptotically flat
spacetimes \cite{hod} where $\omega_r$ approaches a constant while
$\omega_i$ tends to infinity in the limit $n\rightarrow \infty$. In
asymptotically flat spacetimes, the constant $\omega_r$ was claimed
as just the right one to make loop quantum gravity give the correct
result for the black hole entropy with some (not clear yet)
correspondence between classical and quantum states. However such
correspondence seems do not hold in AdS space.

For the large black hole regime the frequencies become evenly spaced
for high overtone number $n$. For lowly charged RNAdS black hole,
choosing bigger values of the charge, the real part in the spacing
expression becomes smaller, while the imaginary part becomes bigger.
This calls for further understanding from CFT.

Qualitative correspondences between quasinormal modes in AdS spaces
and the decay of perturbations in the due CFT have been obtained
above, while quantitive examinations are difficult to be done in
usual four-dimensional black holes due to the mathematical
complicacy. Encouragingly, in the three-dimensional (3D) BTZ black
hole model \cite{BTZ}, the mathematical simplicity can help us to
understand the problem much better. A precise quantitative agreement
between the quasinormal frequencies and the location of poles of the
retarded correlation function of the corresponding perturbations in
the dual CFT has been presented \cite{Bir}. This gives a further
evidence of the correspondence between gravity in AdS spacetime and
quantum field theory at the boundary.

There has been an increasing interest in gravity on de Sitter (dS)
spacetimes in view of recent observational support for a positive
cosmological constant. A holographic duality relating quantum
gravity on D-dimensional dS space to CFT on (D-1)-sphere has been
proposed \cite{Stro}. It is of interest to extend the study in
\cite{Bir} to dS space by displaying the exact solution of the
quasinormal mode problem in the dS bulk space and exploring its
relation to the CFT theory at the boundary. This could serve as a
quantitative test of the dS/CFT correspondence. We have used the
nontrivial 3D dS spacetimes as a testing ground to examine the
dS/CFT correspondence \cite{du1}\cite{du2}. The mathematical
simplicity in these models renders all computations analytical.

The metric of the 3D rotating dS spacetime is given by
\begin{widetext}
\begin{equation} \label{eq(1)}
\begin{split}
ds^{2}=-(M-\frac{{r^2}}{{l^2}}+\frac{{J^2}}{{4r^2}})dt^2+(M-\frac{{r^2}}{{l^2}}+
\frac{{J^2}}{{4r^2}})^{-1}dr^2+r^2(d\varphi-\frac{{J}}{{2r^2}}dt)^2,
\end{split}
\end{equation}
\end{widetext}
where $J$ is associated to the angular momentum. The horizon of such
spacetime can be obtained from
\begin{equation}   \label{eq(2)}
M-\frac{{r^2}}{{l^2}}+\frac{{J^2}}{{4r^2}}=0.
\end{equation}
The solution is given in terms of $r_+$ and $-ir_-$, where $r_+$
corresponds to the cosmological horizon and $-ir_-$ here being
imaginary, has no physical interpretation in terms of a horizon.
Using $r_+$ and $r_-$, the mass and angular momentum of spacetime
can be expressed as
\begin{equation}  \label{eq(3)}
M=\frac{{r_+^2-r_-^2}}{{l^2}}, \hspace{1cm} J=\frac{{-2r_+r_-}}{{l}}
\end{equation}

Scalar perturbations of this spacetime are described by the wave
equation
\begin{equation} \label{eq(4)} \frac{{1}}{{\sqrt{-g}}}\partial
_{\mu }( \sqrt{-g}g^{\mu \nu }\partial _{\nu }\Phi) -\mu^2\Phi =0,
\end{equation}
where $\mu $ is the mass of the field. Adopting the separation
\begin{equation} \label{eq(5)} \Phi (t,r,\varphi )=R(r)\ e^{-i\omega t}\
e^{im\varphi },
\end{equation}
the radial part of the wave equation can be written as
\begin{eqnarray} \label{eq(6)}
\frac{{1}}{{g_{rr}}}\mu^2 R &=& \frac{{1}}{{g_{rr}}}
\frac{{d}}{{rdr}}(\frac{{r^{2}}}{{g_{rr}}}\frac{{dR}}{{rdr}}) +[
\omega ^{2}-\frac{{1}}{{r^{2}}}m^{2}( M- \nonumber \\
&& \frac{{r^{2}}}{{l^{2}}}) -\frac{{J}}{{r^{2}}}m\omega] R,
\end{eqnarray}
where $g_{rr}=(M-r^2/l^2+J^2/(4r^2))^{-1}$. Employing (\ref{eq(3)})
and defining $z=\frac{{r^2-r_+^2}}{{r^2-(-ir_-)^2}}$, the radial
wave equation can be simplified into
\begin{widetext}
\be{ \label{eq(7)}
\begin{split}
(1-z)\frac{{d}}{{dz}}( z\frac{{dR}}{{dz}}) +\left[
\frac{{1}}{{z}}\left( \frac{\omega
l^{2}r_{+}+mlr_{-}}{2(r_{+}^{2}+r_{-}^{2})}\right) ^{2}-\left(
\frac{-\omega l^{2}ir_{-}+imlr_{+}}{2(r_{+}^{2}+r_{-}^{2})} \right)
^{2}+\frac{1}{4(1-z)}\mu^2l^2 \right] R=0.
\end{split}
}
\end{widetext}
We now set the Ansatz \begin{equation}    \label{eq(8)}
R(z)=z^{\alpha }(1-z)^{\beta }F(z),
\end{equation} and Eq.(\ref{eq(7)}) can be transformed into
\begin{widetext}
\begin{equation} \label{eq(9)}
\begin{split}
z(1 - z)\frac{d^{2}F}{dz^{2}} & +  \left[ 1+2\alpha -(1+2\alpha +2\beta )z%
\right] \frac{dF}{dz} +  \{(\beta (\beta -1)+\frac{\mu^2 l^2
}{4})\frac{1}{1-z}+\frac{1}{z}\left[ \left( \frac{\omega
l^{2}r_{+}+mlr_{-}}{2(r_{+}^{2}+r_{-}^{2})}\right) ^{2}+\alpha
^{2}\right] \\
& -  \left[ \left( \frac{-i\omega l^{2}r_{-}+imlr_{+}}{%
2(r_{+}^{2}+r_{-}^{2})}\right) ^{2}+\alpha ^{2}+(1+2\alpha )\beta
+\beta (\beta -1)\right] \}F =0.
\end{split}
\end{equation}
\end{widetext}
Comparing with the standard hypergeometric equation
\begin{equation}
\label{eq(10)}
z(1-z)\frac{d^{2}F}{dz^{2}}+[c-(1+a+b)z]\frac{dF}{dz}-abF=0,
\end{equation}
we have
\begin{eqnarray}    \label{eq(11)}
c & = & 1+2\alpha   \nonumber \\
a+b & = & 2\alpha +2\beta  \nonumber \\
\alpha ^{2}+\left( \frac{\omega l^{2}r_{+}+mlr_{-}}{2(r_{+}^{2}+r_{-}^{2})}%
\right) ^{2} & = & 0 \nonumber \\
\beta (\beta -1)+\frac{\mu^2l^2 }{4} &= &0  \nonumber \\
\left( \frac{-\omega
l^{2}ir_{-}+imlr_{+}}{2(r_{+}^{2}+r_{-}^{2})}\right) ^{2}+(\alpha
+\beta )^{2}. & = & ab
\end{eqnarray}
Without loss of generality, we can take
\begin{eqnarray}
\label{eq(12)}
\alpha &=&-i\left( \frac{\omega l^{2}r_{+}+mlr_{-}}{2(r_{+}^{2}+r_{-}^{2})}%
\right) , \nonumber \\
\beta &=&\frac{1}{2}\left( 1-\sqrt{1-\mu^2 l^2 }\right),
\end{eqnarray}
which leads to
\begin{eqnarray}     \label{eq(13)} a & = &
-\frac{i}{2}\left( \frac{\omega
l^{2}+iml}{r_{+}+ir_{-}}+i(1-\sqrt{1-\mu^2 l^2  })\right),  \nonumber \\
b & = & -\frac{i}{2}\left( \frac{\omega
l^{2}-iml}{r_{+}-ir_{-}}+i(1-\sqrt{1-\mu^2 l^2  })\right), \nonumber \\
c & = & 1-i\left( \frac{\omega
l^{2}r_{+}+mlr_{-}}{r_{+}^{2}+r_{-}^{2}}\right),
\end{eqnarray}
and the solution of (\ref{eq(7)}) reads
\begin{equation}
\label{eq(14)} R(z)=z^{\alpha }(1-z)^{\beta }\ _{2}F_{_{1}}(a,b,c,z) 
\quad .
\end{equation}

Using basic properties of the hypergeometric equation we write the
result as
\begin{widetext}
\be{ \label{eq(15)}
\begin{split}
R(z) &=z^{\alpha }(1-z)^{\beta }(1-z)^{c-a-b}\frac{\Gamma (c)\Gamma (a+b-c)%
}{\Gamma (a)\Gamma (b)}\ _{2}F_{_{1}}(c-a,c-b,c-a-b+1,1-z)  \\
& +z^{\alpha }(1-z)^{\beta }\frac{\Gamma (c)\Gamma (c-a-b)}{\Gamma
(c-a)\Gamma (c-b)}\ _{2}F_{_{1}}(a,b,a+b-c+1,1-z)\quad .
\end{split}
}
\end{widetext}
The first term vanishes at $z=1$, while the second  vanishes
provided that
\begin{equation}       \label{eq(16)} c-a=-n,\qquad
or\qquad c-b=-n,
\end{equation}
where $n=0,1,2,..$. Employing Eqs (\ref{eq(13)}), it is easy to see
that the quasinormal frequencies are
\begin{eqnarray} \label{eq(17)}
\omega _{R} &=&i\frac{m}{l}-2i\left( \frac{r_{+}-ir_{-}}{l^{2}}\right) (n+%
\frac{h_+}{2})  \nonumber \\
\omega _{L} &=&-i\frac{m}{l}-2i\left( \frac{r_{+}+ir_{-}}{l^{2}}\right) (n+%
\frac{h_+}{2}),
\end{eqnarray}
where $h_{\pm}=1\pm\sqrt{1-\mu^2 l^2}$. Taking other values of
$\alpha$ and $\beta$ satisfying (\ref{eq(11)}), we have also the
frequencies
\begin{eqnarray}
\omega _{R} &=&i\frac{m}{l}+2i\left( \frac{r_{+}-ir_{-}}{l^{2}}\right) (n+\frac{h_+}{2})  \nonumber \\
\omega _{L} &=&-i\frac{m}{l}+2i\left(
\frac{r_{+}+ir_{-}}{l^{2}}\right) (n+\frac{h_+}{2}),
\end{eqnarray}
\begin{eqnarray}
\omega _{R} &=&i\frac{m}{l}-2i\left( \frac{r_{+}-ir_{-}}{l^{2}}\right) (n+\frac{h_-}{2})  \nonumber \\
\omega _{L} &=&-i\frac{m}{l}-2i\left(
\frac{r_{+}+ir_{-}}{l^{2}}\right) (n+\frac{h_-}{2}),
\end{eqnarray}
\begin{eqnarray}
\omega _{R} &=&i\frac{m}{l}+2i\left( \frac{r_{+}-ir_{-}}{l^{2}}\right) (n+\frac{h_-}{2})  \nonumber \\
\omega _{L} &=&-i\frac{m}{l}+2i\left(
\frac{r_{+}+ir_{-}}{l^{2}}\right) (n+\frac{h_-}{2}). \label{eq(17a)}
\end{eqnarray}

For a thermodynamical system the relaxation process of a small
perturbation is determined by the poles, in the momentum
representation, of the retarded correlation function of the
perturbation. Let's now investigate the quasinormal modes from the
CFT side. Define the invariant distance between two points defined
by $x$ and $x'$ reads \cite{Stro}\cite{Klemm}
\begin{equation}
\label{eq(20)} d=l\arccos P,
\end{equation}
where $P=X^{A} \eta_{AB} X'^{B} $. In the limit $r, r' \rightarrow
\infty$,
\begin{eqnarray} \label{eq(21)}
P && \approx 2\sinh \frac{{(ir_++r_-)(l\Delta\varphi-i\Delta
t)}}{{2l^2}} \nonumber \\
&&\times \sinh\frac{{(ir_+-r_-)(l\Delta\varphi+i\Delta t)}}{{2l^2}}
\quad .
\end{eqnarray}
This means that we can find the Hadamard Green's function as defined
by \cite{Klemm} in terms of $P$. Such a Green's function is defined
as $G(u,u')=<0\vert\left\{\phi(u),\phi(u')\right\}\vert 0>$ with
$(\nabla_x^2-\mu^2)G=0$. It is possible to obtain the solution
\begin{equation} \label{eq(22)} G\sim F(h_+, h_-, 3/2, (1+P)/2)
\end{equation}
in the limit $r, r'\rightarrow\infty$.

Following \cite{Stro} \cite{Klemm}, we choose boundary conditions
for the fields such that
\begin{equation}
\lim_{r\rightarrow\infty}\phi(r,t,\varphi) \rightarrow
r^{-h_-}\phi_-(t,\varphi)\quad .
\end{equation}
Then, for large $r,r'$, an expression for the two point function of
a given operator $O$ coupling to $\phi$ has the form
\begin{widetext}
\be{ \label{eq(23)}
\begin{split}
\int dt d\varphi dt' d\varphi'\frac{{(rr')^2}}{{l^2}}\phi
\stackrel{\leftrightarrow}{\partial_{r*}} G
\stackrel{\leftrightarrow}{\partial_{r*}} \phi  =  \int dt d\varphi
dt' d\varphi' \phi
\frac{{1}}{{[2\sinh\frac{{(ir_++r_-)(l\Delta\varphi-i\Delta
t)}}{{2l^2}}\sinh\frac{{{(ir_+-r_-)(l\Delta\varphi+i\Delta
t)}}}{{{2l^2}}}]^{h_+}}}\phi
\end{split}
}
\end{widetext}
where $r*$ in (\ref{eq(23)}) is the tortoise coordinate.

For quasinormal modes, we have
\begin{widetext}
\be{ \label{eq(24)}
\begin{split}
& \int dt d\varphi dt' d\varphi'\frac{{exp(-im'\varphi'-i\omega
't'+im\varphi+i\omega
t)}}{{[2\sinh\frac{{(ir_++r_-)(l\Delta\varphi-i\Delta
t)}}{{2l^2}}\sinh\frac{{(ir_+-r_-)(l\Delta\varphi+i\Delta
t)}}{{2l^2}}]^{h_+}}} \\
& \approx
\delta_{mm'}\delta(\omega-\omega')\Gamma(h_+/2+\frac{{im/2l+\omega/2}}{{2\pi
T}}) \Gamma(h_+/2-\frac{{im/2l+\omega/2}}{{2\pi T}})
\Gamma(h_+/2+\frac{{im/2l-\omega/2}}{{2\pi
\bar{T}}})\\
& \times\Gamma(h_+/2-\frac{{im/2l-\omega/2}}{{2\pi \bar{T}}})
\end{split}
}
\end{widetext}
where we changed variables to $v=l\varphi+it, \bar{v}=l\varphi-it$,
and $T=\frac{{ir_+-r_-}}{{2\pi l^2}},
\bar{T}=\frac{{ir_++r_-}}{{2\pi l^2}}$. The poles of such a
correlator are
\begin{eqnarray} \omega_L & = & -\frac{{im}}{{l}}\pm
2\frac{{ir_+ - r_-}}{{l^2}}(n+h_+/2), \\ \nonumber \omega_R & = &
\frac{{im}}{{l}} \pm 2\frac{{ir_+ +r_-}}{{l^2}}(n+h_+/2),
\end{eqnarray}
corresponding to the quasinormal modes (\ref{eq(17)},\ref{eq(17a)})
obtained before.

The quasinormal eigenfunctions thus correspond to excitation of the
corresponding CFT, being exactly those that appear in the spectrum
of the two point functions of CFT operators in dS background for
large values of $r$, that is at the boundary.

The poles of such a correlator correspond exactly to the QNM's
obtained from the wave equation in the bulk. This provides a
quantitative test of the dS/CFT correspondence. This work has been
extended to four-dimensional dS spacetimes  \cite{du2}.

\section{Conclusions and outlooks}

Perturbations around black holes have been a hot topic in general
relativity in the last decades. The main reason is its astrophysical
importance of the study in order to foresee gravitational waves. To
extract as much information as possible from the future
gravitational wave observations, the accurate quasinormal modes
waveforms are needed for different kinds of black holes, including
different stationary black holes (especially interesting with
rotations), time-dependent black hole spacetimes which can describe
the black hole absorption and evaporation processes and extremely
interesting situations with colliding black holes since they may
produce stronger gravitational wave signals which may be easier to be
detected.

Besides astronomical interest, perturbations around black holes can
also serve as a testing ground to examine the recent theories
proposed in string theory, such as the relation between physics in
(A)dS spaces and Conformal Field Theory on its boundary. Very
recently, it was even argued that quasinormal modes could be a way
to detect extra dimensions. String theory makes the radial
prediction that spacetime has extra dimensions and gravity
propagates in higher dimensions. Using the black string model as an
example, it was shown that different from the late time
signal-simple power-law tail in the usual four-dimensions, high
frequency signal persists in the black string \cite{11}. These
frequencies are characters of the massive modes contributed from
extra dimensions. This possible way to detect extra-dimensions needs
further theoretical examinations and it is expected that future
gravitational wave observation could help to prove the existence of
the extra dimensions so that can give support to the fundamental
string theory.

\begin{acknowledgments}
This work  was supported by NNSF of China, Ministry of Education of
China, Shanghai Science and Technology Commission and Shanghai
Education Commission. We acknowledge many helpful discussions with
C. Y. Lin, C. Molina and E. Abdalla.

\end{acknowledgments}

\end{document}